\documentclass{PoS}
\usepackage{slashed}
\usepackage{comment}
\usepackage{graphicx,graphics}
\usepackage{setspace}

\newcommand{\cL}{{\cal L}}

\newcommand{\cO}{{\cal O}}

\newcommand{\ra}{\rightarrow}
\newcommand{\be}{\begin{equation}}
\newcommand{\ee}{\end{equation}}
\newcommand{\bea}{\begin{eqnarray}}
\newcommand{\eea}{\end{eqnarray}}
\newcommand{\Ra}{\Rightarrow}

\newcommand{\baa}{\begin{array}}
\newcommand{\eaa}{\end{array}}
\long\def\symbolfootnote[#1]#2{\begingroup
\def\thefootnote{\fnsymbol{footnote}}\footnote[#1]{#2}\endgroup}
\newcommand{\laf}{\lambda_\phi}
\newcommand{\lam}{\lambda_m}
\newcommand{\las}{\lambda_\sigma}

\title{One-loop potential with 
  scale invariance and effective operators}

\ShortTitle{One-loop scale invariant potential and  effective operators}

\author{\speaker{D. M.  Ghilencea}
\\

CERN Theory Division, CH-1211 Geneva 23, Switzerland and
Theoretical Physics Department, National Institute of Physics and Nuclear Engineering (IFIN-HH) Bucharest,
 077125 Romania

  E-mail: \email{dumitru.ghilencea@cern.ch}
}

\abstract{
We study quantum corrections to the scalar potential in classically
scale invariant theories, using a  manifestly scale invariant regularization.
To this purpose, the subtraction scale $\mu$ of the dimensional regularization 
 is generated by spontaneous scale symmetry breaking, from a  subtraction 
{\it function} of the fields,   $\mu(\phi,\sigma)$. This  function is then
uniquely determined from  general principles showing that it depends on 
the dilaton only, with  $\mu(\sigma)\sim \sigma$.
The result is a scale invariant one-loop potential $U$ for a higgs
field $\phi$ and dilaton $\sigma$ that  contains an  additional {\it finite}
 quantum correction $\Delta U(\phi,\sigma)$,  beyond the Coleman Weinberg term.
$\Delta U$ contains new,  non-polynomial effective operators like $\phi^6/\sigma^2$
whose quantum  origin is explained.  A flat direction is maintained at the quantum level, the  model 
has vanishing vacuum energy and  the one-loop correction to the mass of  
$\phi$ remains small without tuning (of its self-coupling, etc) beyond the  initial, classical
tuning (of the dilaton coupling) that enforces a hierarchy $\langle\sigma\rangle\gg \langle\phi\rangle$.  
The approach is useful to models that investigate 
scale symmetry at the quantum level.}

\FullConference{Proceedings of the \\
Corfu Summer Institute 2015 "School and Workshops on Elementary Particle Physics and Gravity"\\
		1-27 September 2015, Corfu, Greece}

\begin{document}

\section{General considerations}

Scale invariant theories \cite{W} were often considered as an alternative to 
supersymmetry to  address the hierarchy problem. Since such theories forbid 
the presence of dimensionful parameters in the Lagrangian, scale symmetry
be it classical or quantum,  must be broken  in the real world.
This breaking  can be explicit or spontaneous.  We investigate the
latter case, since this  preserves the UV behaviour of the initial scale invariant theory.
In the spontaneous breaking  we have the dilaton $\sigma$ as the Goldstone mode of this symmetry,
with non-vanishing vacuum expectation value (vev)  $\langle\sigma\rangle\not\!=\!0$.
In a broader setup that includes gravity,  $\langle\sigma\rangle$ may be related to the Planck 
scale. We do not  detail how $\sigma$ acquires a vev, but simply search for solutions 
with $\langle\sigma\rangle\not\!=\!0$. All scales of the theory are then 
related  to  $\langle\sigma\rangle$. A hierarchy of scales which are
vev's of the  different fields present, can then  be generated either by
 {\it one}  initial (classical) tuning of the couplings to 
small  values  \cite{GGR} or as in  \cite{KA}.

The purpose of this talk  based on \cite{dmg}  is to consider a classically scale invariant
theory and to show how to  compute the quantum 
corrections to the scalar potential  in a manifestly scale invariant way. 
Such approach is important since it preserves at the quantum level the initial 
symmetry of the theory and its UV properties, relevant for the Higgs physics.
Investigating scale invariant theories at quantum level is non-trivial
 because the regularization  of  their  quantum corrections 
breaks the scale symmetry {\it explicitly}\footnote{
As a result of such explicit breaking the  initial flat direction 
of the classically scale invariant theory is  lifted; there is extensive model 
building in this direction, see for example more recent 
\cite{K,Y,AF,H,B,G,EG,I,I2,Khoze,Khoze2,K2}.}.
Indeed, in  its  traditional form, the  regularization, be it dimensional regularization (DR)
or some other scheme, introduces a subtraction  scale\footnote{In DR the  scale $\mu$ relates 
the dimensionless (renormalized) coupling $\lambda^{(r)}$  to the  dimensionful one $\lambda$
  once the $d=4$ theory is continued analytically to $d\!=\!4\!-\!2\epsilon$;
for a quartic higgs coupling: 
$\lambda = \mu^{2\epsilon} \,\big(\lambda^{(r)}+\sum_n a_n/\epsilon^n\big)$. }
$\mu$. 
Its presence ruins exactly the symmetry that we  want to study at the quantum level. 
To avoid this situation one should 
 generate the subtraction scale in a dynamical way.
Consider then replacing $\mu$ of the DR scheme  by a field-dependent subtraction {\it function} 
$\mu(\sigma)$ \cite{Englert}, see also more recent \cite{S1}.
 Having couplings or masses that are field-dependent is something familiar in string theory. 
After spontaneous breaking of scale symmetry when $\langle\sigma\rangle\not=0$, 
the subtraction scale is generated as $\mu(\langle\sigma\rangle)$.
Doing so has implications at the quantum level, presented below.

Preserving the  scale symmetry of the action   during  the UV regularization of the
quantum correction is actually required in theories which are non-renormalizable, 
to avoid regularization  artefacts. Since some of these theories may be
 be non-renormalizable  \cite{S2,GM},  it is then worth exploring the 
consequence of a such regularization.
This is also important for the naturalness problem, as  argued
long ago by Bardeen  \cite{Bardeen}. The Standard Model (SM) with the higgs mass $m_h=0$ has an
extra symmetry, classical scale invariance; therefore, if one uses schemes that break 
this  symmetry, which is  what happens in general,  quadratic divergences
can be regarded as artefacts  of this regularization and can thus be ignored\footnote{
$m_{h}^2$ remains quadratic  in the scale (of ``new physics'')  generated by 
spontaneous scale symmetry breaking.}. 
  This gives a phenomenological motivation to study a scale invariant regularization
of quantum corrections, with spontaneous breaking of scale symmetry.   
This  talk (based on \cite{dmg}) continues previous similar studies  \cite{S1,S2,S3}, 
with some notable differences and with  new results shown below.

We consider a classical, scale invariant theory
of a  Higgs-like scalar $\phi$ and the dilaton $\sigma$, with 
\bea\label{init}
\cL=\frac{1}{2}\,\partial_\mu \phi\, \partial^\mu \phi
+\frac{1}{2}\,\partial_\mu \sigma\, \partial^\mu \sigma
-V(\phi,\sigma)
\eea
where 
\bea\label{v}
V=\frac{\lambda_\phi}{4}\phi^4+\frac{\lambda_m}{2}\,\phi^2\sigma^2
+\frac{\lambda_\sigma}{4}\,\sigma^4
\eea

\medskip\noindent
There exists a non-trivial solution  with spontaneous breaking 
of scale symmetry   $\langle\sigma\rangle\not=0$,
(also $\langle\phi\rangle\not=0$) provided that two conditions are met:
firstly,
 $\lambda_m^2=\lambda_\phi\lambda_\sigma$ with $\lambda_m<0$ and secondly,
\medskip
\bea\label{vv}
\frac{\langle\phi\rangle^2}{\langle \sigma\rangle^2}
 =-\frac{\lambda_m}{\lambda_\phi},
\quad\Ra\quad
V=\frac{\lambda_\phi}{4}\Big(\,
\phi^2+\frac{\lambda_m}{\lambda_\phi}\,\sigma^2\,\Big)^2
\eea

\medskip\noindent
Then spontaneous scale symmetry breaking implies electroweak symmetry breaking
at tree-level, with $V_{min}=0$ i.e. vacuum energy is vanishing. There is a flat direction
corresponding to a massless dilaton $\sigma$ while the mass of $\phi$ is
$m_{\tilde\phi}^2
=- 2\lambda_m\,(1-\lambda_m/\lambda_\phi)\,\langle\sigma\rangle^2$.
Although we are not interested in the exact value of $\langle\sigma\rangle$,
since scale invariance is expected to be broken by Planck physics
it is likely  that $\langle\sigma\rangle\sim M_{Planck}$.
To ensure a hierarchy of scales $\langle\phi\rangle\ll\langle\sigma\rangle$ and  that
the  mass of the higgs-like $\phi$ is near the electroweak scale, 
one can tune {\it once} the couplings (at the classical level) to enforce a 
relation $\lambda_\sigma\ll\vert\lambda_m\vert\ll\lambda_\phi$ \cite{GGR,KA}.
Such hierarchy of couplings is stable under the quantum corrections of the 
renormalization group flow \cite{GGR}. 
 One would like to explore these issues at the quantum level
in a  scale invariant approach.

More generally, in  theories with scale symmetry, the  potential for $\phi$ and $\sigma$
has the form $V=\sigma^4\, W(\phi/\sigma)$. Assuming spontaneous breaking of this symmetry
 $\langle\sigma\rangle\not=0$, the two minimum conditions for $V$ become
\medskip
\bea\label{W2}
W'(x_0)=W(x_0)=0,
\qquad 
x_0\equiv\frac{\langle\phi\rangle}{\langle\sigma\rangle};\,
\quad \langle\sigma\rangle, \langle\phi\rangle\not=0.
\eea 

\medskip\noindent
One minimum condition fixes the ratio $\langle\phi\rangle/\langle\sigma\rangle$ 
while the second gives a relation among the couplings of the theory such as $W(x_0)=0$.
 If these two equations for $W$  have a solution $x_0$,
then $\langle\phi\rangle\propto \langle\sigma\rangle$. 
A flat direction exists
in the plane $(\phi,\sigma)$ with $\phi/\sigma=x_0$ along which the vacuum energy vanishes
$V(\langle\phi\rangle,\langle\sigma\rangle)=0$.
These results can  remain valid at the {\it quantum level} as well, provided that
the calculation of the quantum corrections is manifestly scale invariant, since 
then the potential remains of the form $V\sim \sigma^4 \tilde W(\phi/\sigma)$
where $\tilde W$ is the quantum corrected $W$.
Also, the flat direction corresponding to the 
Goldstone mode (dilaton)  remains flat at the quantum level
(to all orders) if  the calculation  preserves the 
scale symmetry which is broken only spontaneously.

\section{Quantum scale invariance of the potential and effective operators}

Let us then calculate the one-loop potential in a scale invariant regularization.
To enforce dimensionless couplings  in the usual DR scheme,
 one  replaces the quartic couplings $\lambda\ra \lambda\mu^{4-d}$, with
  $\mu$ the subtraction scale.
 This breaks classical  scale symmetry.  To avoid this, 
replace $\mu$ by  an unknown function of the fields, $\mu(\phi,\sigma)$, 
whose vacuum expectation value  generates dynamically the subtraction 
scale\footnote{after spontaneous scale symmetry breaking.} $\mu(\langle\phi\rangle,\langle\sigma\rangle)$.
 This function is determined later, but is assumed to be continuous and differentiable
at all fields values. Then, in $d=4-2\epsilon$ dimensions, the potential is
\bea\label{VV}
\tilde V(\phi,\sigma)=
\mu(\phi,\sigma)^{4-d}\,V(\phi,\sigma)
\eea
%
To be general, we also allowed a $\phi$-dependence of the subtraction function. 
There are now ``evanescent'' interactions 
between say $\sigma$ in $\mu(\phi,\sigma)$ and $V(\phi,\sigma)$, that are absent 
in the limit $d=4$. They are due to the  scale symmetry in $d=4-2\epsilon$.
The one-loop potential is then computed as usual, but with $\tilde V$ instead of
original $V$:
 \medskip
\bea
U&=&\tilde V(\phi,\sigma) -
\frac{i}{2}\,
\,\int \frac{d^d p}{(2\pi)^d}
\,{\rm Tr}\ln \big[ p^2-\tilde V_{\alpha\beta}+i\varepsilon\big]
\eea

\medskip\noindent
Here  $\tilde V_{\alpha\beta}=\partial \tilde V/\partial\alpha\partial\beta$, 
with $\alpha,\beta=\phi,\sigma$. Up to $\cO(\epsilon^2)$ terms
\bea\label{m2}
\tilde V_{\alpha\beta}
&=&
\mu^{2\epsilon}\,\Big[V_{\alpha\beta} + 2\epsilon \,\mu^{-2}\,N_{\alpha\beta}\Big],
\qquad
\\[6pt]
N_{\alpha\beta}&\equiv &\mu\, (\mu_\alpha\, V_\beta +\mu_\beta\, V_\alpha)
 +(\mu\, \mu_{\alpha\beta}  -\mu_\alpha\,\mu_\beta)\,V,
\label{m2prime}
\eea
 
 \medskip\noindent
where $\mu_\alpha={\partial\mu}/{\partial\alpha}$, $\mu_{\alpha\beta}
={\partial^2 \mu}/{\partial\alpha\partial\beta}$, $V_\alpha=\partial V/\partial\alpha$, 
and $V_{\alpha\beta}=\partial^2 V/\partial\alpha\partial\beta$,
are field dependent quantities. Denote by $M_s^2$  the eigenvalues of the
matrix $V_{\alpha\beta}$ and\footnote{The eigenvalues $M_s^2$ are the roots $(q)$ of
$q^2-q\,(V_{\phi\phi}+V_{\sigma\sigma})+(V_{\phi\phi} V_{\sigma\sigma}-V_{\phi\sigma}^2)=0$.}
by $\kappa\equiv 4\pi e^{3/2-\gamma_E}$,  then 
\medskip
\bea\label{lt}
U 
=\mu(\phi,\sigma)^{2\epsilon}
\Big\{
V- \frac{1}{64 \pi^2}\,
\Big[
 \sum_{s=\phi,\sigma} 
M^4_s \,  \Big(\,  \frac{1}{\epsilon}
- \ln\frac{M_s^2(\phi,\sigma)}{\kappa\,\mu^2(\phi,\sigma)} \Big)
+
\frac{4 \,(V_{\alpha\beta}\,N_{\beta\alpha})}{\mu^2(\phi,\sigma)}\Big]\Big\}
\eea

\medskip\noindent
In the last term a summation over repeated indices (fields) is understood.
The counterterms are
\medskip
\bea\label{Z}
 \delta U_{c.t.}\equiv \mu(\phi,\sigma)^{2\epsilon} 
\Big\{ \frac14\, \delta Z_{\lambda_\phi}\,\, \lambda_\phi\,\phi^4+
 \frac12\,\delta Z_{\lambda_m}\,\,\lambda_m\,\phi^2\,\sigma^2
+\frac14 \,\delta Z_{\lambda_\sigma}\,\,\lambda_\sigma\,\sigma^4\big]
\eea 

\medskip\noindent
from which one easily finds the coefficients $\delta Z_{\lambda_j}\equiv Z_{\lambda_j}-1$; they
have values identical to those  if the theory were regularized 
with $\mu$=constant (the same is true about the beta functions of $\lambda_j$).
For example $\delta Z_{\lambda_\phi}=1/(16\pi^2\epsilon)(9\lambda_\phi+\lambda_m^2/\lambda_\phi)$, etc.
The renormalized potential is then
\bea\label{U}
U(\phi,\sigma)&=&V(\phi,\sigma)+\frac{1}{64\pi^2}\,
\Big\{
\sum_{s=\phi,\sigma} M^4_s(\phi,\sigma)\, \Big(\ln \frac{M^2_s(\phi,\sigma)}{\mu^2(\phi,\sigma)}
-\frac{3}{2}\,\Big)
+\Delta U(\phi,\sigma)
\Big\}
\nonumber\\[4pt]
\Delta U
\!\!&=&\!\! \frac{-4}{\mu^2} \Big\{
V\,\big[ (\mu\mu_{\phi\phi}-\mu_\phi^2)\,V_{\phi\phi}
+2\,(\mu\mu_{\phi\sigma}-\mu_\phi\mu_\sigma) V_{\phi\sigma}
+(\mu\mu_{\sigma\sigma}-\mu_\sigma^2)\,V_{\sigma\sigma}\big]
\nonumber\\[4pt]
&+&
2\mu\,(\mu_\phi\,V_{\phi\phi}+\mu_\sigma\,V_{\phi\sigma})\,V_\phi
+2\mu\,(\mu_\phi\,V_{\phi\sigma}+\mu_\sigma\,V_{\sigma\sigma})\,V_\sigma\Big\}
\eea

\medskip\noindent
We recovered the usual Coleman-Weinberg term \cite{C1,G1} in a modified, scale invariant form. 
We also found  a new, additional contribution  $\Delta U$, not considered in \cite{S1},  which  
is a  {\it finite} one-loop correction to the potential. 
 $\Delta U$ emerges from the correction $\propto\epsilon$ to the field dependent masses
(eq.(\ref{m2})) of the states that ``run'' in the loop;
  when this correction $\propto \epsilon$ multiplies the pole $1/\epsilon$ of the 
 momentum integral, one obtains a  finite term (last term in (\ref{lt})) that
gives  $\Delta U$. Also $\Delta U$ vanishes on the tree level ground state
when $V$ and its first derivatives vanish.

The problem with the result in eq.(\ref{U}) is that it 
 depends on the unknown function  $\mu(\phi,\sigma)$, which generates 
the subtraction scale  after spontaneous scale symmetry breaking. 
Obviously,  physical observables  cannot depend on the regularization 
done with this function. We should then  determine this function 
from some general principles; then,  the potential must respect
the Callan-Symanzik equation,  to enable us to make physical predictions.

To this purpose, consider first a particular example of $\mu(\phi,\!\sigma)$
used in previous  models \cite{S1,S3}
\bea\label{muS}
\mu(\phi,\sigma)=z\, \big(\xi_\phi\,\phi^2+\xi_\sigma\,\sigma^2\big)^{1/2}
\eea
With this, one computes the expression of  $\Delta U$ which in the particular
limit $\lambda_m\!\ra\! 0$ becomes
\medskip
\bea
\Delta U\Big\vert_{\lambda_m=0}&=&  
 -3 \,
\Big[\xi_\phi\xi_\sigma\,\big[\lambda_\phi\, (9\lambda_\phi+\lambda_\sigma)\,\phi^6\sigma^2
+\lambda_\sigma (\lambda_\phi+9\lambda_\sigma)\,\phi^2\sigma^6 \big]
\nonumber\\[5pt]
&&\,\, +\,\, 7\,\big(\lambda_\phi^2 \,\xi_\phi^2\,\phi^8
+\lambda_\sigma^2\,\xi_\sigma^2\,\sigma^8\big)\,
-
 (\xi_\phi^2+\xi_\sigma^2)
 \lambda_\phi\lambda_\sigma\,\phi^4\sigma^4 \Big](\xi_\phi\phi^2+\xi_\sigma\sigma^2)^{-2}
\label{nondec}
\eea

\medskip\noindent
This simplifies further if  $\lambda_m^2=\lambda_\phi\lambda_\sigma$ which ensures
 spontaneous breaking of scale symmetry;
 however the term $\propto \xi_\phi\xi_\sigma \lambda_\phi^2\phi^6\sigma^2$ remains even in this case,
unless $\xi_\phi\xi_\sigma=0$ when $\mu$ depends on one field only.
 Now,  when $\lambda_m\ra 0$, the ``visible'' sector of higgs-like $\phi$ is classically decoupled
from the ``hidden'' sector of the dilaton $\sigma$. Nevertheless, we see that in 
this limit the two sectors still interact at the quantum level, which is unacceptable.
This situation is more general and applies when the subtraction function depends on
both fields. The reason for this is related to how $\Delta U$ is generated, 
from ``evanescent'' interactions  introduced by scale invariance of the 
action in $d=4-2\epsilon$, see  $\tilde V(\phi,\sigma)=\mu^{2\epsilon} V(\phi,\sigma)$.
Since $\mu(\phi,\sigma)$ contains both fields,  it brings interactions 
with any term in $V(\phi,\sigma)$, not only with that proportional to $\lambda_m$. 
This explains the presence in (\ref{nondec}) of non-decoupling interactions terms
proportional to $\lambda_\phi$. 

Similar considerations  apply for a general subtraction function, which, up to a 
relabeling of the fields, can be  written as
$\mu(\phi,\sigma)=z \,\sigma \exp( g(\phi/\sigma))$, with $g$ an arbitrary 
continuous, differentiable function. Then
one can show that $\Delta U$ vanishes in the decoupling limit ($\lambda_m=0$)  only
if $g$ is a constant\footnote{We discard  an extra solution for $g$ and thus $\mu(\phi,\sigma)$
which  is  not continuous in $\phi=0$ and also  depends on 2 arbitrary 
constants  rather than  one ($z$), where the latter is taken care of by the Callan-Symanzik equation,
(see later).}. We conclude that the subtraction  function must depend on the dilaton
 only, with  $\mu(\sigma)= z\sigma$. Here $z$ is some arbitrary dimensionless parameter,
whose role will be clarified shortly\footnote{To be exact,
one actually has $\mu(\sigma)=z\sigma^{2/(d-2)}$, since the fields have mass dimension $(d-2)/2$
while $\mu$ has mass dimension 1. In our one-loop approximation and at this stage it is safe
to take the limit $d\ra 4$ in which case $\mu(\sigma)=z \sigma$.}.

With $\mu(\sigma)=z\sigma$ uniquely identified and with  $V$ of eq.(\ref{v}) one  obtains
\medskip
\be\label{tr}
\Delta U =
\frac{\laf\lam \phi^6}{\sigma^2}
-
\big(16 \laf\lam+6\lam^2-3 \laf\las\big)\phi^4
-\big(16\lam+25\las\big)\,\lambda_m\,\phi^2\sigma^2
- 21\las^2\sigma^4
\ee

\medskip\noindent
This simplifies further for our  case with $\lambda_m^2=\lambda_\phi\lambda_\sigma$
of spontaneous symmetry breaking
that generates a subtraction scale $z\langle\sigma\rangle$.
In the decoupling limit ($\lambda_m\ra 0$) there are no
quantum interactions left between $\phi$ and $\sigma$, since $\Delta U\!\ra\! 0$.
With $\Delta U$ of eq.(\ref{tr}) the one-loop potential becomes
\medskip
\bea\label{fi}
U(\phi,\sigma)=
V(\phi,\sigma)
+\frac{1}{64\pi^2}\,
\Big[
\sum_{s=\phi,\sigma} M^4_s(\phi,\sigma)\, \Big(\ln \frac{M^2_s(\phi,\sigma)}{z^2 \sigma^2}
-\frac{3}{2}\,\Big)
+\Delta U(\phi,\sigma)\Big]
\eea

\medskip\noindent
This quantum expression  is  scale invariant. It has a form 
and properties similar to those discussed in section~1 (text around eq.(\ref{W2})).

\section{More about quantum corrections}

One notes  the presence in $\Delta U$ of eq.(\ref{tr})
of higher dimensional {\it non-polynomial} operators
such as $\phi^6/\sigma^2$,  in addition to other finite quantum results induced by
manifest scale invariance. It is expected that more such operators be generated
at higher loop orders. 
Needless to say, the correction $\Delta U$ is missed in calculations that 
are not  scale invariant such as the usual DR scheme, since the result depends 
on derivatives  of $\mu$ wrt $\sigma$ which vanish if $\mu$=constant.
Finally, after a Taylor expansion, the above  operator can be re-written 
as a series of standard effective polynomial operators in fluctuations 
$\tilde\sigma$, where  $\sigma=\langle\sigma\rangle+\tilde\sigma$  
\bea
\frac{\phi^6}{\sigma^2}=\frac{\phi^6}{\langle\sigma\rangle^2}
\Big(1-\frac{2\tilde\sigma}{\langle\sigma\rangle}
+3\frac{\tilde\sigma^2}{\langle\sigma\rangle^2}+\cdots\Big)
\eea
The logarithm $\ln (z^2\sigma^2)$ of the Coleman-Weinberg (CW) term can also be expanded
about $\langle\sigma\rangle$ to recover the usual CW term  obtained for 
$\mu$=constant ($=z\langle\sigma\rangle$), plus additional corrections.
In conclusion, $U$ contains new quantum corrections that can be re-written
as series of polynomial terms in $\tilde\sigma$, suppressed
by $\langle\sigma\rangle$.

At higher loops, the new operator $\phi^6/\sigma^2$ can render the theory  
non-renormalizable.  If  the initial theory were regularized 
with $\mu$=constant, the scale symmetry is broken explicitly,
there is no Goldstone mode, but such operator is never  generated dynamically 
and the theory is renormalizable to all orders. Nevertheless such
operators should be added ``by hand'' to the classical Lagrangian, since they
respect its symmetries. The presence here of this non-polynomial operator is not too
surprising;
since it is not  forbidden by the scale symmetry (preserved by the quantum
calculation)
such operator is expected to  emerge  at some loop-order. Its origin
is due to loop corrections with ``evanescent'' interactions 
dictated by scale  invariance in  $d=4-2\epsilon$.

The potential  $U$ still depends on the dimensionless 
subtraction parameter $z$, which apparently prevents one from making 
predictions for  physical observables. 
However, its presence is  understood
by analogy to the subtraction {\it scale} dependence (in a given order) 
in the ``ordinary'' regularization with $\mu$=constant.
The Callan-Symanzik equation should be respected by the
potential and this will  ensure that the physical observables do not
depend on $z$ (or on the subtraction scale  $\mu(\langle\sigma\rangle)=
z\langle\sigma\rangle$). In our case the Callan-Symanzik  equation for
$U$ of eq.(\ref{fi}) is
\smallskip
\bea
\frac{d U(\lambda_j,z)}{d\ln z}=\Big(\beta_{\lambda_j}\,\frac{\partial }{\partial \lambda_j}
+\frac{\partial}{\partial z}\Big) U(\lambda_j,z)= 0, 
\qquad\rm{sum\,\,over}\,\, j=\phi,m,\sigma; 
\quad 
\beta_{\lambda_j}\equiv \frac{d\alpha_j}{d\ln z}
\eea

\smallskip\noindent
This condition is easily verified, since
 the only explicit dependence on $z$ is via the CW term
and the (non-vanishing \cite{tamarit,AMS}) beta functions\footnote{
These are
$\beta_{\lambda_\phi}=1/(8\pi^2) \,(9\lambda_\phi^2+\lambda_m^2)$,
$\beta_{\lambda_m}=1/(8\pi^2) \,
(3\lambda_\phi+4\lambda_m+3\lambda_\sigma)\lambda_m$ and
$\beta_{\lambda_\sigma}=1/(8\pi^2) \,(\lambda_m^2+9\lambda_\sigma^2)$.}
$\beta_{\lambda_j}$ are found from the condition
$(d / d\ln z) (\lambda_j Z_{\lambda_j})=0$  ($j=\phi,m, \sigma$, fixed).
In conclusion the one-loop potential is independent of $z$ and respects
the Callan-Symanzik equation for theories with this symmetry \cite{tamarit}.

Since $U$ is scale invariant  at the one-loop level,
the necessary minimum conditions of vanishing first derivatives
 $U_\phi=U_\sigma=0$, 
with $\langle\sigma\rangle\not=0$, ensure that the ground state has
vanishing vacuum energy $U_{min}=0$ and that a flat direction exists
(as discussed in Section~1). The spectrum consists of a massless
(Goldstone) mode and the scalar $\phi$ receives quantum corrections.
Its mass is then
$m_\phi^2=(U_{\phi\phi}+U_{\sigma\sigma})_{min}$.
One can compute $m_\phi$ in some approximation, such as
$\lambda_\sigma\ll\vert\lambda_m\vert\ll \lambda_\phi$, when minimising the potential
is easier.  It is possible to  show that the quantum correction to the mass of $\phi$
due to the Coleman-Weinberg part of the potential does not require additional
tuning of the couplings in order to keep it light \cite{S1}, well below the scale
$\langle\sigma\rangle$, where
$\langle\sigma\rangle\gg\langle\phi\rangle$ (see also text after eq.(\ref{vv})). 
This means  that there are no quantum corrections to its mass 
of the type $\lambda_\phi^2\langle\sigma\rangle^2$ or similar, that would require tuning
the higgs quartic self-coupling  $\lambda_\phi$ \footnote{
Such tuning of $\lambda_\phi$  would be the sign of re-introducing
the hierarchy problem in the context discussed here.}.

Further, the contribution to the (mass)$^2$ of $\phi$ due to the new correction $\Delta U$ that
we found (not considered in \cite{S1}) is also under control. This correction is
\smallskip
\bea
\delta m_\phi^2&=&
\frac{1}{64\pi^2}\,(\Delta U_{\phi\phi}+\Delta U_{\sigma\sigma})_{min}
\nonumber
\eea

\smallskip\noindent
It can be shown \cite{dmg} that $\delta m_\phi^2$ contains only terms proportional to
 $\lambda_m^2\langle\sigma\rangle$ or  $\lambda_\sigma\langle\sigma\rangle$  
(here $\lambda_\sigma\!=\!\lambda_m^2/\lambda_\phi\!\ll\!\vert\lambda_m\vert$). Therefore no tuning of 
the higgs self-coupling $\lambda_\phi$ is required to keep $\delta m_\phi^2$  much  smaller than 
the UV scale $\langle\sigma\rangle$.

\section{Final remarks and conclusions}

Scale invariant theories can provide an alternative to supersymmetry to
address the mass hierarchy problem. 
The  Standard Model classical Lagrangian  has a scale symmetry in 
the limit of vanishing tree-level higgs mass. As emphasized long ago
by Bardeen,  the usual regularization  of  quantum corrections breaks 
this symmetry  {\it explicitly} by the presence of the subtraction scale 
(via dimensional  regularization, Pauli-Villars, etc)
and introduces regularization artefacts.
Obviously,  in the real world scale symmetry is broken,  but to preserve its
 ultraviolet properties, while generating all the mass scales of the theory 
(including the subtraction scale), it is
recommended that  this symmetry be broken only  spontaneously (softly).  
This means the spectrum of the theory will contain an additional, massless  
(Goldstone) mode  of  this symmetry  (dilaton $\sigma$), whose vev generates the 
subtraction scale. All other scales, vacuum expectations of scalar fields, 
are then related to $\langle\sigma\rangle$.

In this talk we presented the consequences of such an approach, with a 
regularization that preserves scale symmetry, to  compute the one-loop corrections
 to the scalar potential  of a classically scale invariant theory of a higgs-like 
 $\phi$ and dilaton $\sigma$.
One consequence is that the one-loop scalar potential contains additional {\it finite} 
quantum corrections ($\Delta U$), beyond the familiar Coleman-Weinberg term,
itself modified into a  scale-invariant form (with $\mu(\sigma)=z\sigma$, where $z$ is 
an arbitrary dimensionless parameter).  The origin of $\Delta U$
is due to evanescent corrections (i.e. proportional to $\epsilon$)
to the field dependent masses that ``run'' in the one-loop diagram
when these multiply the pole of its momentum integral, to give a finite $\Delta U$.
Also $\Delta U$ contains new, non-polynomial effective operators of the type $\phi^6/\sigma^2$.
These  can be Taylor expanded into a series of polynomial  operators,
suppressed by $\langle\sigma\rangle\gg\langle\phi\rangle$; note that no dangerous
operators of the opposite type $\sigma^6/\phi^2$ can be generated.
 The quantum correction to the  mass of $\phi$ ($m_\phi$)
 due to $\Delta U$ remains thus under control, with no tuning needed of the higgs self-coupling
to keep $m_\phi$ much smaller than the dilaton scale $\langle\sigma\rangle$.
It would be interesting to check  if this behaviour survives in higher loop orders.

It was also shown that the subtraction function cannot also depend on the higgs-like scalar $\phi$.
This is because in  the classical decoupling limit $\lambda_m\ra 0$
of the visible sector (of $\phi$) from the hidden sector (of $\sigma$), there 
exists a non-decoupling quantum  interaction between these sectors. As a result the subtraction
function depends on $\sigma$ only $\mu(\sigma)=z\sigma$, as considered above, and is {\it  unique}. 
Since physical observables cannot depend on arbitrary parameters such as $z$ (or the
subtraction scale $\mu(\langle\sigma\rangle)=z\langle\sigma\rangle$),
we checked that the Callan-Symanzik equation is respected by the potential.
The above results can now be applied to the scale invariant
version of the (classical) Standard Model Lagrangian to explore their phenomenological
implications. The presence of the higher dimensional operators of the type found above
can have  implications for the stability of the SM ground state at the high
scale.

\vspace{0.5cm}
\noindent
{\bf Acknowledgements:  }
This work  was supported by a grant of the Romanian National Authority for 
Scientific Research CNCS-UEFISCDI,  project number PN-II-ID-PCE-2011-3-0607.


\begin{thebibliography}{90}
\bibitem{W}
For an early work see
  C.~Wetterich,
 ``Cosmology and the Fate of Dilatation Symmetry,''
  Nucl.\ Phys.\ B {\bf 302} (1988) 668.


\bibitem{GGR}
  K.~Allison, C.~T.~Hill and G.~G.~Ross,
  ``Ultra-weak sector, Higgs boson mass, and the dilaton,''
  Phys.\ Lett.\ B {\bf 738} (2014) 191
  [arXiv:1404.6268 [hep-ph]].


\bibitem{KA}
  A.~Kobakhidze,
``Quantum relaxation of the Higgs mass,''
  arXiv:1506.04840 [hep-ph].


\bibitem{dmg}
  D.~M.~Ghilencea,
``Manifestly scale-invariant regularization and quantum effective operators,''
  Phys.~Rev.~D {\bf 93} (2016) no.10,  105006, 
  [arXiv:1508.00595 [hep-ph]].

\bibitem{K}
  R.~Foot, A.~Kobakhidze, K.~L.~McDonald and R.~R.~Volkas,
 ``A Solution to the hierarchy problem from an almost decoupled hidden sector 
within a classically scale invariant theory,''
  Phys.\ Rev.\ D {\bf 77} (2008) 035006
  [arXiv:0709.2750 [hep-ph]].



\bibitem{Y}
  K.~Endo and Y.~Sumino,
  ``A Scale-invariant Higgs Sector and Structure of the Vacuum,''
  JHEP {\bf 1505} (2015) 030
  [arXiv:1503.02819 [hep-ph]].

\bibitem{AF}
  A.~Farzinnia, H.~J.~He and J.~Ren,
  ``Natural Electroweak Symmetry Breaking from Scale Invariant Higgs Mechanism,''
  Phys.\ Lett.\ B {\bf 727} (2013) 141
  [arXiv:1308.0295 [hep-ph]].

\bibitem{H}
  C.~T.~Hill,
  ``Is the Higgs Boson Associated with Coleman-Weinberg Dynamical Symmetry Breaking?,''
  Phys.\ Rev.\ D {\bf 89} (2014) 7,  073003
  [arXiv:1401.4185 [hep-ph]].

\bibitem{B}
  B.~Grinstein and P.~Uttayarat,
``A Very Light Dilaton,''
  JHEP {\bf 1107} (2011) 038
  [arXiv:1105.2370 [hep-ph]].

 \bibitem{G}
  W.~D.~Goldberger, B.~Grinstein and W.~Skiba,
  ``Distinguishing the Higgs boson from the dilaton at the Large Hadron Collider,''
  Phys.\ Rev.\ Lett.\  {\bf 100} (2008) 111802
  [arXiv:0708.1463 [hep-ph]].



\bibitem{EG}
  E.~Gabrielli, M.~Heikinheimo, K.~Kannike, A.~Racioppi, M.~Raidal and C.~Spethmann,
  ``Towards Completing the Standard Model: Vacuum Stability, EWSB and Dark Matter,''
  Phys.\ Rev.\ D {\bf 89} (2014) 1,  015017
  [arXiv:1309.6632 [hep-ph]].

\bibitem{I}
  S.~Iso, N.~Okada and Y.~Orikasa,
``Classically conformal $B^-$ L extended Standard Model,''
  Phys.\ Lett.\ B {\bf 676} (2009) 81
  [arXiv:0902.4050 [hep-ph]].

\bibitem{I2}
  S.~Iso and Y.~Orikasa,
``TeV Scale B-L model with a flat Higgs potential at the Planck scale 
- in view of the hierarchy problem -,''
  PTEP {\bf 2013} (2013) 023B08
  [arXiv:1210.2848 [hep-ph]].

\bibitem{Khoze}
  V.~V.~Khoze,
``Inflation and Dark Matter in the Higgs Portal of Classically Scale Invariant Standard Model,''
  JHEP {\bf 1311} (2013) 215
  [arXiv:1308.6338 [hep-ph]].

\bibitem{Khoze2}
  V.~V.~Khoze, C.~McCabe and G.~Ro,
``Higgs vacuum stability from the dark matter portal,''
  JHEP {\bf 1408} (2014) 026
  [arXiv:1403.4953 [hep-ph]].

 \bibitem{K2}
  R.~Foot, A.~Kobakhidze, K.~L.~McDonald and R.~R.~Volkas,
  ``Poincar\'e  protection for a natural electroweak scale,''
  Phys.\ Rev.\ D {\bf 89} (2014) 11,  115018
  [arXiv:1310.0223 [hep-ph]].


\bibitem{Englert}
  F.~Englert, C.~Truffin and R.~Gastmans,
  ``Conformal Invariance in Quantum Gravity,''
  Nucl.\ Phys.\ B {\bf 117} (1976) 407.
  S.~Deser,
  ``Scale invariance and gravitational coupling,''
  Annals Phys.\  {\bf 59} (1970) 248.


\bibitem{S1}
  M.~Shaposhnikov and D.~Zenhausern,
  ``Quantum scale invariance, cosmological constant and hierarchy problem,''
  Phys.\ Lett.\ B {\bf 671} (2009) 162
  [arXiv:0809.3406 [hep-th]].

\bibitem{S2}
  M.~E.~Shaposhnikov and F.~V.~Tkachov,
``Quantum scale-invariant models as effective field theories,''
  arXiv:0905.4857 [hep-th].


\bibitem{GM}
  F.~Gretsch and A.~Monin,
``Dilaton: Saving Conformal Symmetry,''
  arXiv:1308.3863 [hep-th].

\bibitem{Bardeen}
  W.~A.~Bardeen,
  ``On naturalness in the standard model,''
  FERMILAB-CONF-95-391-T, C95-08-27.3.

 \bibitem{S3}
   M.~Shaposhnikov and D.~Zenhausern,
   ``Scale invariance, unimodular gravity and dark energy,''
   Phys.\ Lett.\ B {\bf 671} (2009) 187
  [arXiv:0809.3395 [hep-th]].

\bibitem{C1}
  S.~R.~Coleman and E.~J.~Weinberg,
  ``Radiative Corrections as the Origin of Spontaneous Symmetry Breaking,''
  Phys.\ Rev.\ D {\bf 7} (1973) 1888.

\bibitem{G1}
  E.~Gildener and S.~Weinberg,
``Symmetry Breaking and Scalar Bosons,''
  Phys.\ Rev.\ D {\bf 13} (1976) 3333.


\bibitem{tamarit}
  C.~Tamarit,
  ``Running couplings with a vanishing scale anomaly,''
  JHEP {\bf 1312} (2013) 098
  [arXiv:1309.0913 [hep-th]]. 

\bibitem{AMS}
  R.~Armillis, A.~Monin and M.~Shaposhnikov,
 ``Spontaneously Broken Conformal Symmetry: Dealing with the Trace Anomaly,''
  JHEP {\bf 1310} (2013) 030
  [arXiv:1302.5619 [hep-th]].
\end{thebibliography}
 \end{document}